\documentclass[aps,prc,showpacs,twocolumn,letter]{revtex4}

\usepackage{graphicx,epsfig}

\begin{document}

\title{Triangular flow in event-by-event ideal hydrodynamics in Au+Au collisions
at $\sqrt{s_{\rm NN}}=200A$ GeV}

\author{Hannah Petersen}
\affiliation{Department of Physics, Duke University, Durham, North Carolina
27708-0305, United States}

\author{Guang-You Qin}
\affiliation{Department of Physics, Duke University, Durham, North Carolina
27708-0305, United States}

\author{Steffen A. Bass}
\affiliation{Department of Physics, Duke University, Durham, North Carolina
27708-0305, United States}

\author{Berndt M\"uller}
\affiliation{Department of Physics, Duke University, Durham, North Carolina
27708-0305, United States}


\begin{abstract}
The first calculation of triangular flow $v_3$ in Au+Au collisions at
$\sqrt{s_{\rm NN}}=200A$ GeV from an
event-by-event (3+1)-d transport+hydrodynamics hybrid approach is presented. As
a response to the initial
triangularity $\epsilon_3$ of the collision zone, $v_3$ is computed in a similar
way to the
standard event-plane analysis for elliptic flow $v_2$.
It is found that the triangular flow exhibits weak centrality dependence and is
roughly equal to elliptic flow in most central collisions.
We also explore the transverse momentum and rapidity dependence of $v_2$ and
$v_3$
for charged particles as well as identified particles. We conclude that an
event-by-event
treatment of the ideal hydrodynamic evolution starting with realistic initial
conditions generates the main features
expected for triangular flow.

\end{abstract}

\keywords{Relativistic Heavy-ion collisions, Monte Carlo simulations,
Hydrodynamic models}

\pacs{25.75.-q,24.10.Lx,24.10.Nz}

\maketitle

Collective behaviour of particles emitted from relativistic heavy ion
collisions, such as elliptic flow, is one of the earliest predicted
observables that indicates
fluid-like behaviour of the created hot and
dense nuclear matter
\cite{Ollitrault:1992bk,Voloshin:1994mz,Kolb:1999it,Teaney:2000cw}. The pressure
gradients
need to be large enough to translate an early stage coordinate space asymmetry
to a final state momentum space
anisotropy. Therefore, the high values of the second coefficient of the Fourier
expansion of the azimuthal
distribution of the particles, $v_2$, that have been reported by the experiments
at the Relativistic Heavy Ion
Collider (RHIC) \cite{Ackermann:2000tr,Adler:2003kt,Adams:2003am,Back:2004zg}
have led to the
conclusion that the quark gluon plasma
is a nearly perfect liquid \cite{Adams:2005dq,Romatschke:2007mq}.

The collective flow observables manifest themselves in
multi-particle-correlations as well. For example, the so called
cumulant method has been very successful in quantifying the harmonic
coefficients of the azimuthal
particle distributions \cite{Borghini:2000sa}. Recently $\Delta
\eta$-$\Delta \phi$ correlations have been explored in a new manner by
extracting
 a triangular flow signal from the data that is
responsible for most of the structures that were previously attributed to other
mechanisms \cite{Alver:2010gr,Alver:2010dn}.
Features like long-range rapidity correlations on the near- and away-side
accompanied by a conical structure on the
away-side have been often referred to in the context of jet-medium interaction.
The preliminary
PHOBOS data show a long range correlation in rapidity which would be supported
by an initial state generated from a
flux tube picture like in NEXspheRIO \cite{Takahashi:2009na,Andrade:2009em}. In
this model one is also able to
observe the features like the ridge and the "cone" in the two-particle
correlations.

The triangular flow, $v_3$, is the third coefficient of the Fourier expansion of
the azimuthal distribution of the
final particles in momentum space with respect to the corresponding event plane
angle $\Psi_3$ that is defined below. This angle fluctuates randomly event by
event in contrast to the well-known event plane angle $\Psi_2$ used for
the elliptic flow analysis, which is strongly correlated to the reaction plane.
The triangular flow is assumed to be the response to a
triangular shape of the initial state, described by a nonzero triangularity, $\epsilon_3$, that arises from the fluctuations of the
initial collisions. 

In contrast to the extensively studied observables like directed flow ($v_1$),
elliptic flow ($v_2$) and the fourth harmonic ($v_4$), the triangular flow
($v_3$) is not correlated to the reaction
plane that is defined by the beam axis and
the impact parameter axis of the collision. The initial state fluctuations in
the transverse plane are random with
respect to the reaction plane. In a standard hydrodynamic calculation with
smooth initial conditions only the even
coefficients of the Fourier expansion are non-zero at midrapidity. The odd
coefficients vanish by symmetry in collisions of identical nuclei which is
the reason why they have not been studied so far.

In this paper the latest version of the Ultra-relativistic Quantum Molecular
Dynamics (UrQMD) \cite{Bass:1998ca,Bleicher:1999xi} together with ideal
relativistic fluid dynamics is used to explore this new observable \footnote{The
code is available as UrQMD-3.3p1 at http://urqmd.org}. The
full event-by-event setup of this
hybrid approach allows to extract a $v_3$ component from the final particle
distributions. The method is very similar to
the standard elliptic flow event plane measurement and will be outlined below.
Predictions for the impact parameter
dependence and the transverse momentum dependence of identified particles are
made.

Let us now review the main ingredients of the hybrid approach
\cite{Petersen:2008dd,Petersen:2009vx} that are relevant
for the development of triangular flow. The initial binary nucleon-nucleon
collisions are modeled in UrQMD following the Lund model of nucleon-nucleon
reactions \cite{Andersson:1983ia} involving color flux tubes excitation
and fragmentation processes that provide long range rapidity correlations and
fluctuations in the energy deposition in
the transverse plane. For Au+Au collisions at the highest RHIC energies the
starting time for the hydrodynamic
evolution has been chosen to be $t_{\rm start}=0.5$ fm in order to fit the final pion
multiplicity at midrapidity. Only the
matter around midrapidity ($|y|<2$) is considered to be locally thermalized and
takes part in the ideal hydrodynamic
evolution. The more dilute spectator/corona regions are treated in the hadronic
cascade approach throughout the reaction. To map the point
particles from the UrQMD initial state to energy, momentum and net baryon
density distributions each particle is
represented by a three-dimensional Gaussian distribution
\cite{Steinheimer:2007iy}.

The ideal hydrodynamic evolution \cite{Rischke:1995ir,Rischke:1995mt} for the
hot and dense stage of the collision
translates the initial fluctuations in the transverse energy density to momentum
space distributions. A hadron gas
equation of state \cite{Zschiesche:2002zr} has been used because we are aiming
here
only at qualitative statements and
not at quantitative comparisons. This equation of state (EoS) has been
extensively tested and gives
reasonable results for multiplicities and
particle spectra. Furthermore, given the same average speed of sound during the
evolution, flow observables are not sensitive to the details of the EoS
\cite{Steinheimer:2009nn,Petersen:2010md}.

The transition from the hydrodynamic evolution to the transport approach when
the matter is diluted in the late stage
is treated as a gradual transition on an approximated constant proper time 
hyper-surface (see
\cite{Li:2008qm,Steinheimer:2009nn} for details). The final rescatterings and
resonance decays are taken into account
in the hadronic cascade.

The above event-by-event setup includes all the main ingredients that are
supposed to be necessary for the build up of
triangular flow. Since the complete final state particle distributions are
calculated, an analysis similar to those applied by experimentalists is used.
%

\begin{figure}[h]
\resizebox{0.5\textwidth}{!}{ \centering
\includegraphics{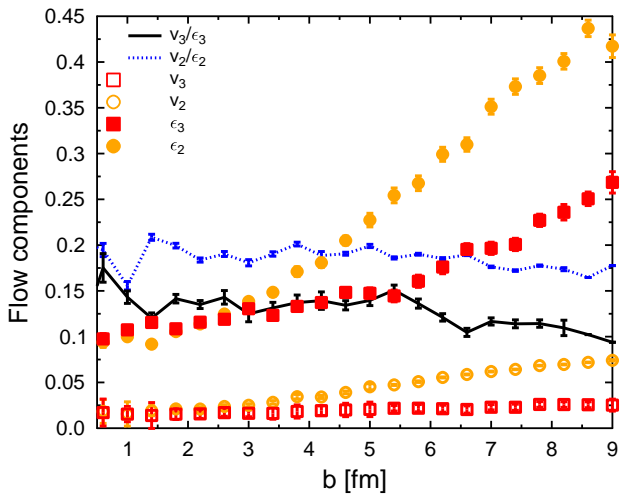}
} 
\vspace{-1cm}
\caption{(Color online) Impact parameter dependence of $v_3$ (open squares)
and $v_2$ (open circles) of charged
particles in Au+Au collisions at $\sqrt{s_{\rm NN}}=200A$ GeV at midrapidity
($|\eta|<1$). The full and the
dotted line represent the ratios of
$v_3/\epsilon_3$ and $v_2/\epsilon_2$ respectively.} \label{fig_v3_v2_imp}
\end{figure}

The definition of the participant eccentricity can be generalized to the
triangularity defined as
\begin{equation}
\epsilon_n=\frac{\sqrt{\langle r^n \cos(n \phi)\rangle^2+\langle r^n \sin(n
\phi)\rangle^2}}{\langle r^n \rangle}
\end{equation}
where in contrast to \cite{Alver:2010gr} the factor in front of $\cos(3\phi)$ is
taken to be $r^3$ since this is the
more consistent way to do the Fourier expansion of the distribution function.
$(r,\phi)$ in
this case are polar coordinates corresponding to the transverse plane in
coordinate space. We have calculated the triangularity of the
UrQMD initial state for all the particles that are newly
produced or have undergone
at least one interaction and with a
rapidity between $-2<y<2$ at the starting time of the hydrodynamic evolution
($t_{\rm start}=0.5$ fm). The centrality dependence and the values of the
eccentricity as well as the triangularity are comparable to those obtained by
a Glauber Monte Carlo approach \cite{Alver:2010dn} (see Fig.
\ref{fig_v3_v2_imp}).

The particle distribution in coordinate space is then transferred to the final state particle distribution in momentum spaceby the
pressure gradients during the hydrodynamic
evolution.
Experiments are only able to measure the momenta
of the particles but not the coordinates, therefore, one has to find a way to
generalize the elliptic flow analysis to triangular flow analysis in a consistent way. We propose here to use the
standard event plane method
\cite{Poskanzer:1998yz} and define an event plane for triangular flow in the
following way
\begin{equation}
\label{eqn_defpsi}
\Psi_n=\frac{1}{n} {\rm arctan}\frac{\langle p_T \sin(n\phi_p)\rangle}{\langle
p_T \cos(n\phi_p)\rangle} \quad ,
\end{equation}
where $(p_T,\phi_p)$ are polar coordinates in momentum space. 

\begin{figure}[h]
\vspace{-.5cm}
\resizebox{0.5\textwidth}{!}{ \centering
\includegraphics{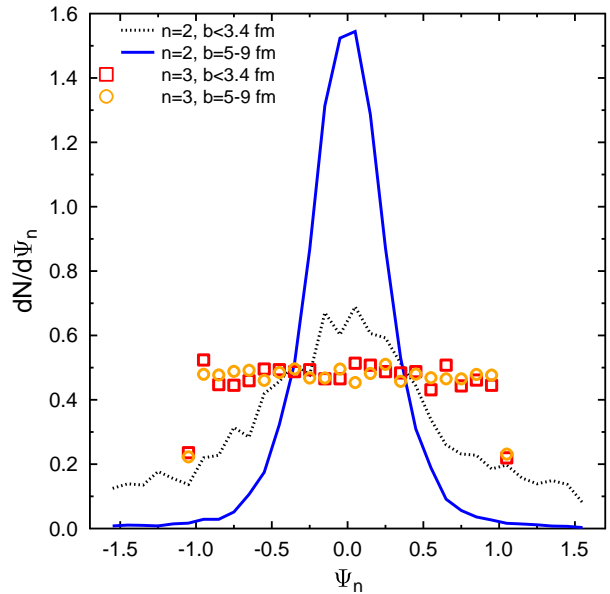}
}
\vspace{-1cm}
\caption{(Color online) Distribution of event-plane angles $\Psi_2$ (dotted
and full line) and $\Psi_3$ (open
squares and circles) with respect to the reaction plane in central ($b<3.4$ fm)
and mid-central ($b=5-9$ fm) Au+Au
collisions at $\sqrt{s_{\rm NN}}=200A$ GeV.} \label{fig_psi_dist}
\vspace{-.5cm}
\end{figure}

As all the details of the final particles are known in our calculation, the
resolution of the event plane angle can be improved by
taking into account all the particles (also neutral particles) in
a certain kinematic range ($|\eta|<2$) to
determine the event plane. The resolution is calculated by comparing two sub events
and turns out to be around 0.8 radians
for the triangular flow plane angle and 0.7 to 0.95 for the $v_2$ event plane angle 
depending on the centrality. This corresponds
to an uncertainty in the angle of approximately 12 degrees in mid-central
collisions.

The distributions of the resulting event plane angles $\Psi_2$ and $\Psi_3$ with
respect to the known reaction plane
(positive x-direction defines $\Psi=0$) are shown in Fig. \ref{fig_psi_dist}. As expected the
elliptic flow event plane is Gaussian
distributed and therefore correlated to the reaction plane, especially in less
central events. The
triangular flow plane shows a flat distribution between -60 and +60 degrees
since the fluctuations that lead to a
$v_3$ component are random with respect to the reaction plane. These internally
consistent results increase our
confidence in the analysis method proposed here.

To explore the correlation between the initial spatial event plane and the
corresponding final event plane we define an initial
event plane angle $\Phi_n$ in analogy to Eq. (\ref{eqn_defpsi}): 
\begin{equation}
\Phi_n=\frac{1}{n} {\rm arctan}\frac{\langle r^n \sin(n\phi)\rangle}{\langle r^n
\cos(n\phi)\rangle} \quad .
\end{equation}
For elliptic flow this angle is the one that defines the so called participant
plane. With the used conventions $\Phi_n$ is defined in the region between
$-\pi/n$ and $+\pi/n$. For convenience, we introduce a shifted angle
$\Phi_n^\prime=\Phi_n+\pi/n$ [mod $2\pi/n$], defined in the domain $-\pi/n<\Phi^\prime_n<+\pi/n$.

\begin{figure}[h]
\vspace{-0.5cm}
\resizebox{0.5\textwidth}{!}{ \centering
\includegraphics{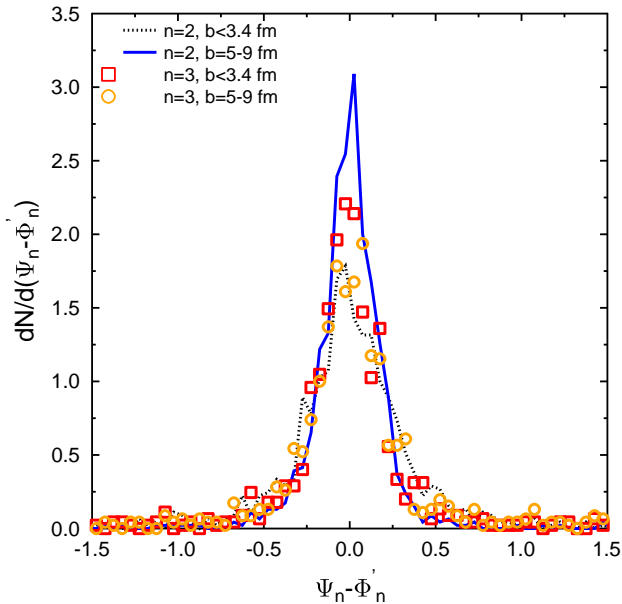}
} 
\vspace{-1cm}
\caption{(Color online) Correlation of the final event plane angles $\Psi_2$
(dotted
and full line) and $\Psi_3$ (open
squares and circles) with the corresponding initial event plane angles $\Phi_2^\prime$
and $\Phi_3^\prime$ in central ($b<3.4$ fm)
and mid-central ($b=5-9$ fm) Au+Au
collisions at $\sqrt{s_{\rm NN}}=200A$ GeV.} \label{fig_phi_psi_corr}
\end{figure}

The correlation between the final ($\Psi_n$) and initial ($\Phi_n^\prime$) event planes
for two different centrality classes is shown in Fig. \ref{fig_phi_psi_corr}.
There is a strong correlation in all four cases which has a similar
shape. For elliptic flow there is a stronger contribution
from the collision geometry that results in a centrality dependence of the correlation ($\Psi_2-\Phi_2^\prime$)
\cite{Qin:2010pf}. In more central collisions most of the elliptic flow comes
from fluctuations while in more peripheral events the almond-shape geometry of
the collision zone has a major effect on $v_2$ which leads to an even stronger
correlation of the two angles. For the
triangular flow the distribution is very similar for central and mid-central
events since it is only caused by fluctuations in both cases. This analysis has been carried
out on an event by event basis and confirms that the final triangular flow
is related to the initial triangularity.  

\begin{figure}[h]
\resizebox{0.5\textwidth}{!}{ \centering
\includegraphics{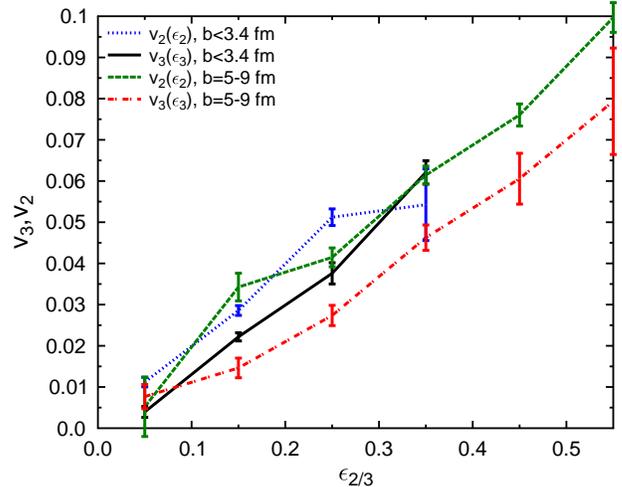}
} 
\vspace{-1cm}
\caption{(Color online) Dependence of $v_2$ and $v_3$ on the initial
$\epsilon_2$ and $\epsilon_3$ of
charged particles in central ($b<3.4$
fm) and mid-central ($b=5-9$ fm) Au+Au collisions at $\sqrt{s_{\rm NN}}=200A$
GeV at midrapidity ($|\eta|<1$). } \label{fig_v_eps_corr}
\end{figure}

Further evidence for a strong correlation between initial state geometry and the
final state momentum anisotropies is found by looking at the flow coefficients
as a function of the corresponding $\epsilon_n$. 
The flow coefficients are calculated by using the following formula
\begin{equation}
v_n=\langle \cos(n(\phi_p-\Psi_n))\rangle
\end{equation}
where it is important to note that the particle that is correlated to the event
plane is removed from the event plane
determination to eliminate auto-correlations. The final results for $v_n$ are
obtained by dividing the above values by
the event plane resolution of the corresponding centrality class. The same
procedure has been applied in the
event-by-event approach of Holopainen et al. \cite{Holopainen:2010gz}.

In Fig. \ref{fig_v_eps_corr} we show $v_2(\epsilon_2)$ and $v_3(\epsilon_3)$ for two different centrality classes ($b<3.4$ fm and $b=5-9$ fm). All the curves
behave linearly within error bars, a larger initial eccentricity/triangularity
leads to a larger elliptic/triangular flow. For $v_3$ the translation of the
initial anisotropy into the final anisotropy is less efficient for non-central
collisions than for central collisions. For elliptic flow the lines have a
similar slope, so the elliptic flow response is similar in central and more
peripheral collisions.


The impact parameter dependence of the flow
coefficients for charged particles is shown in Fig. \ref{fig_v3_v2_imp}.
First of all, the triangular flow has a finite value which exhibits only a weak
centrality dependence. This is another hint that
$v_3$ is only induced by fluctuations in contrast to $v_2$ which has a geometry
influence in addition that is centrality dependent. $\epsilon_2$ and $\epsilon_3$ increase as a function of
impact parameter since the almond shape of the overlap region gets more
pronounced in peripheral collisions and the fluctuations are more important in
smaller systems. For the third Fourier coefficient the ratio of $v_3/\epsilon_3$
decreases slightly for more peripheral collisions which reflects the shorter duration
of the hydrodynamic evolution. There is less time to translate the initial
state anisotropy to a final state momentum anisotropy.
Furthermore, the relative elliptic
flow is larger than the relative triangular flow, so the transfer of coordinate
space anisotropy to momentum space
anisotropy is more efficient for lower harmonics than for higher ones. The same
result has already been found for
$v_4$ that is much smaller than $v_2$ \cite{:2010ux}.

\begin{figure}[h]
\resizebox{0.5\textwidth}{!}{ \centering
\includegraphics{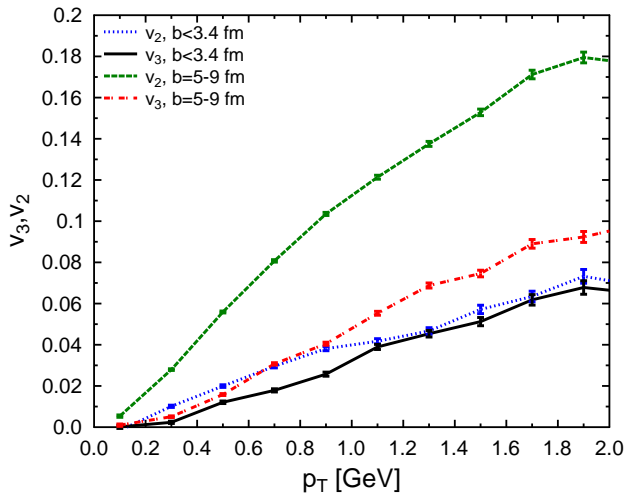}
} 
\vspace{-1cm}
\caption{(Color online) Transverse momentum dependence of $v_3$ and $v_2$ of
charged particles in central ($b<3.4$
fm) and mid-central ($b=5-9$ fm) Au+Au collisions at $\sqrt{s_{\rm NN}}=200A$
GeV at midrapidity ($|\eta|<1$).} \label{fig_pt_dep}
\end{figure}

The momentum
dependence of $v_3$ and $v_2$ for charged particles is shown in Fig.
\ref{fig_pt_dep}. It shows that the elliptic flow is almost twice as large as the triangular flow in mid-central collisions. The elliptic
flow results extracted with the event
plane method are nonzero even in the most central collisions due to the
fluctuations of the participant plane with
respect to the reaction plane.
In central collisions the magnitude and the shape of the triangular flow is
similar to
the elliptic flow. This can be traced back to the fact the distribution of spatial
event plane angles $P(\Psi_2)$ and $P(\Psi_3)$ are flat in central collisions
(b=0 fm), rendering no correlation
between $v_2$/$v_3$ and the reaction plane. The remaining correlation for
central collisions (see Fig. \ref{fig_psi_dist}) of $\Psi_2$ with the reaction
plane arises because there is a contribution from finite impact parameter
calculations ($b<3.4$ fm) in the most central class of events. 

\begin{figure}[h]
\resizebox{0.5\textwidth}{!}{ \centering
\includegraphics{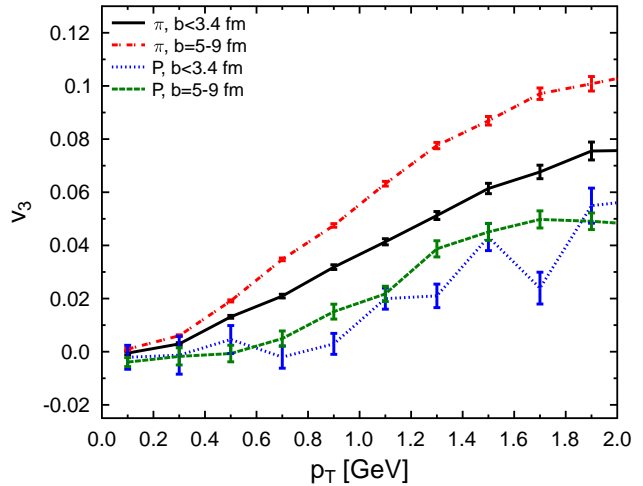}
} 
\vspace{-1cm}
\caption{(Color online) Transverse momentum dependence of $v_3$ for pions and
protons in central ($b<3.4$ fm) and
mid-central ($b=5-9$ fm) Au+Au collisions at $\sqrt{s_{\rm NN}}=200A$ GeV at
midrapidity ($|\eta|<1$).}
\label{fig_ptid_dep}
\end{figure}

The transverse momentum dependence of
triangular flow for identified particles is shown in Fig. \ref{fig_ptid_dep}.
We have checked that the elliptic flow for identified particles is compatible
with the
experimental data. The fireball created at the highest RHIC energies is
dominated by mesons and thus the pion flow
is very similar to the charged particle flow. For
the protons, the same mass
splitting effect is seen for $v_3$ as for elliptic flow \cite{Petersen:2010md}.
In addition, the proton $v_3$ is almost equal for central and mid-central
collisions.

\begin{figure}[h]
\resizebox{0.5\textwidth}{!}{ \centering
\includegraphics{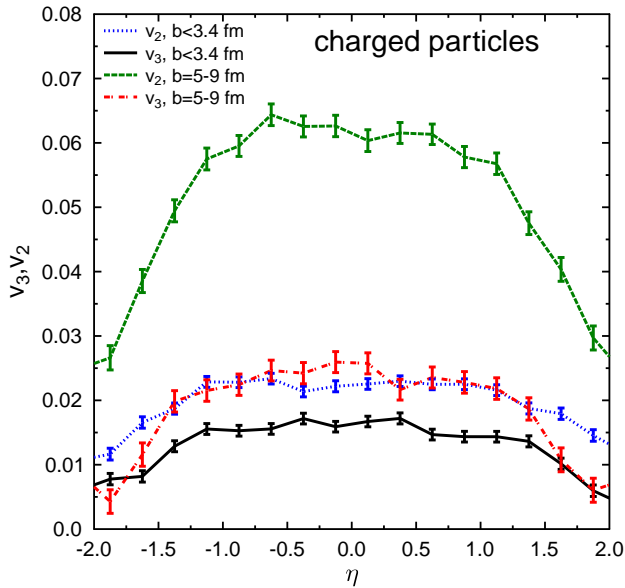}
} 
\caption{(Color online) Pseudorapidity dependence of $v_3$ and $v_2$ for
charged particles in Au+Au collisions at
$\sqrt{s_{\rm NN}}=200A$ GeV.} \label{fig_eta_dep}
\end{figure}

Fig. \ref{fig_eta_dep} shows the pseudorapidity dependence of the two flow
coefficients for charged particles in two
centrality classes. Due to the initial conditions generated by UrQMD with its reliance on fluxtube fragmentation one obtains a long-range $\Delta \eta$
correlation that can be observed in the final
state.
The elliptic flow results are flat for at least two units of pseudorapidity
whereas the triangular flow
distribution is almost flat over the whole pseudorapidity range ($\Delta \eta=4$) covered by the present calculation.

In conclusion, we have presented the first calculation of triangular flow from a
(3+1)d ideal
hydrodynamics approach in Au+Au collisions
at $\sqrt{s_{\rm NN}}=200A$ GeV. The fluctuating initial
conditions and the event-by-event setup
are crucial for this observable. By translating initial state triangularity to
the final state momentum distributions
via pressure gradients a finite value of the third coefficient of the Fourier
expansion of the azimuthal distribution
of the particles in the final state is generated.

Our method is based on a generalization of the standard event plane analysis
that has been used for elliptic
flow and can therefore be done in experiments in exactly the same way. While
$v_2$ shows a strong impact parameter
dependence, $v_3$ exhibits only a weak centrality dependence and is close to $v_2$ in
central collisions.
The transverse momentum dependence of $v_3$ is similar to the elliptic flow and
also the mass splitting is observed for identified particles.
The flat rapidity dependence that results from the color flux tubes in the
initial conditions is in agreement with the observation of the ridge in
$\Delta \eta$-$\Delta \phi$ correlations.
By measuring $v_3$ also in a differential way one might be able to learn
something about the amount and the size of the initial state fluctuations.

\section*{Acknowledgements}
\label{ack} We are grateful to the Open Science Grid for the computing
resources. The authors thank Dirk Rischke for
providing the 1 fluid hydrodynamics code. H.P. acknowledges a Feodor Lynen
fellowship of the Alexander von Humboldt
foundation. This work was supported in part by U.S. department of Energy grant
DE-FG02-05ER41367 and NSF grant PHY-09-41373. H. Petersen thanks
Burak Alver and Michael Mitrovski for fruitful discussions. The authors thank
Jean-Yves Ollitrault for his comments to the manuscript. Furthermore, the INT
Seattle is acknowledged for support
to participate in the program "Quantifying the properties of Hot and Dense QCD
matter" where the idea for this
publication was born.



\begin{thebibliography}{99}

\bibitem{Ollitrault:1992bk}
  J.~Y.~Ollitrault,
  Phys.\ Rev.\  D {\bf 46}, 229 (1992).

\bibitem{Voloshin:1994mz}
  S.~Voloshin and Y.~Zhang,
  Z.\ Phys.\  C {\bf 70}, 665 (1996).

\bibitem{Kolb:1999it}
  P.~F.~Kolb, J.~Sollfrank and U.~W.~Heinz,
  Phys.\ Lett.\  B {\bf 459}, 667 (1999).

\bibitem{Teaney:2000cw}
  D.~Teaney, J.~Lauret and E.~V.~Shuryak,
  Phys.\ Rev.\ Lett.\  {\bf 86}, 4783 (2001).

\bibitem{Ackermann:2000tr}
  K.~H.~Ackermann {\it et al.}  [STAR Collaboration],
  Phys.\ Rev.\ Lett.\  {\bf 86}, 402 (2001).

\bibitem{Adler:2003kt}
  S.~S.~Adler {\it et al.}  [PHENIX Collaboration],
  Phys.\ Rev.\ Lett.\  {\bf 91}, 182301 (2003).

\bibitem{Adams:2003am}
  J.~Adams {\it et al.}  [STAR Collaboration],
  Phys.\ Rev.\ Lett.\  {\bf 92}, 052302 (2004).

\bibitem{Back:2004zg}
  B.~B.~Back {\it et al.}  [PHOBOS Collaboration],
  Phys.\ Rev.\ Lett.\  {\bf 94}, 122303 (2005).

\bibitem{Adams:2005dq}
  J.~Adams {\it et al.}  [STAR Collaboration],
  Nucl.\ Phys.\  A {\bf 757}, 102 (2005).

\bibitem{Romatschke:2007mq}
  P.~Romatschke and U.~Romatschke,
  Phys.\ Rev.\ Lett.\  {\bf 99}, 172301 (2007).

\bibitem{Borghini:2000sa}
  N.~Borghini, P.~M.~Dinh and J.~Y.~Ollitrault,
  Phys.\ Rev.\  C {\bf 63}, 054906 (2001).

\bibitem{Alver:2010gr}
  B.~Alver and G.~Roland,
  Phys.\ Rev.\  C {\bf 81}, 054905 (2010).

\bibitem{Alver:2010dn}
  B.~H.~Alver, C.~Gombeaud, M.~Luzum and J.~Y.~Ollitrault,
  arXiv:1007.5469 [nucl-th].


\bibitem{Takahashi:2009na}
  J.~Takahashi {\it et al.},
  Phys.\ Rev.\ Lett.\  {\bf 103}, 242301 (2009).

\bibitem{Andrade:2009em}
  R.~Andrade, F.~Grassi, Y.~Hama and W.~L.~Qian,
  arXiv:0912.0703 [nucl-th].
\bibitem{Bass:1998ca}
  S.~A.~Bass {\it et al.},
  Prog.\ Part.\ Nucl.\ Phys.\  {\bf 41}, 255 (1998)
  [Prog.\ Part.\ Nucl.\ Phys.\  {\bf 41}, 225 (1998)].
\bibitem{Bleicher:1999xi}
  M.~Bleicher {\it et al.},
  J.\ Phys.\ G {\bf 25}, 1859 (1999).

\bibitem{Petersen:2008dd}
  H.~Petersen, J.~Steinheimer, G.~Burau, M.~Bleicher and H.~Stocker,
  Phys.\ Rev.\  C {\bf 78}, 044901 (2008).

\bibitem{Petersen:2009vx}
  H.~Petersen and M.~Bleicher,
  Phys.\ Rev.\  C {\bf 79}, 054904 (2009).

\bibitem{Andersson:1983ia}
  B.~Andersson, G.~Gustafson, G.~Ingelman and T.~Sjostrand,
  Phys.\ Rept.\  {\bf 97}, 31 (1983).

\bibitem{Steinheimer:2007iy}
  J.~Steinheimer, M.~Bleicher, H.~Petersen, S.~Schramm, H.~Stocker and
D.~Zschiesche,
  Phys.\ Rev.\  C {\bf 77}, 034901 (2008).

\bibitem{Rischke:1995ir}
  D.~H.~Rischke, S.~Bernard and J.~A.~Maruhn,
  Nucl.\ Phys.\  A {\bf 595}, 346 (1995).

\bibitem{Rischke:1995mt}
  D.~H.~Rischke, Y.~Pursun and J.~A.~Maruhn,
  Nucl.\ Phys.\  A {\bf 595}, 383 (1995)
  [Erratum-ibid.\  A {\bf 596}, 717 (1996)].

\bibitem{Zschiesche:2002zr}
  D.~Zschiesche, S.~Schramm, J.~Schaffner-Bielich, H.~Stoecker and W.~Greiner,
  Phys.\ Lett.\  B {\bf 547}, 7 (2002).

\bibitem{Steinheimer:2009nn}
  J.~Steinheimer, V.~Dexheimer, H.~Petersen, M.~Bleicher, S.~Schramm and
H.~Stoecker,
  Phys.\ Rev.\ C {\bf 81}, 044913 (2010).

\bibitem{Petersen:2010md}
  H.~Petersen and M.~Bleicher,
  Phys.\ Rev.\  C {\bf 81}, 044906 (2010).

\bibitem{Li:2008qm}
  Q.~f.~Li, J.~Steinheimer, H.~Petersen, M.~Bleicher and H.~Stocker,
  Phys.\ Lett.\  B {\bf 674}, 111 (2009).

\bibitem{Poskanzer:1998yz}
  A.~M.~Poskanzer and S.~A.~Voloshin,
  Phys.\ Rev.\  C {\bf 58}, 1671 (1998).
\bibitem{Qin:2010pf}
  G.~Y.~Qin, H.~Petersen, S.~A.~Bass and B.~Muller,
  arXiv:1009.1847 [nucl-th].

\bibitem{Holopainen:2010gz}
  H.~Holopainen, H.~Niemi and K.~J.~Eskola,
  arXiv:1007.0368 [hep-ph].

\bibitem{:2010ux}
   and A.~Adare  [The PHENIX Collaboration],
  arXiv:1003.5586 [nucl-ex].



\end{thebibliography}
\end{document}